\documentclass{article}

\usepackage{graphicx}
\usepackage{amsmath}
\usepackage[noadjust]{cite}

\begin{document}

\baselineskip 18pt

\begin{center}
{\Large {\bf  Quantum Confinement Induced Shift in Energy Band Edges and Band Gap of Spherical Quantum Dot}}

\vskip5mm P. Borah$^{1}$, D. Siboh$^{1}$, P. K. Kalita$^{1}$, J. K. Sarma$^{2}$ and N. M. Nath $^{1}$\footnote{Corresponding author, E-mail: nayanmaninath@gmail.com} \vskip3mm \mbox{}%

$^{1}$ Department of Physics, Rajiv Gandhi University, Rono Hills, Doimukh-791112, Arunachal pradesh, India

\mbox{}$^{2}$ Department of Physics, Tezpur University, Tezpur-784028, Assam, India

\bigskip

\begin{abstract}
We have proposed and validated an ansatz as effective potential for confining electron/hole within spherical quantum dot in order to understand quantum confinement and its consequences associated with energy states and band gap of Spherical Quantum Dot. Within effective mass approximation formalism, considering an ansatz incorporating a conjoined harmonic oscillator and coulomb interaction as the effective potential for confining an electron or a hole within a spherical quantum dot and by employing appropriate boundary conditions we have calculated the shifts in energy of minimum of conduction band(CBM) and maximum of valence band(VBM) with respect to size of spherical quantum dot. We have also determined the quantum confinement induced shift in band gap energy of spherical quantum dot. In order to verify our theoretical predictions as well as to validate our ansatz, we have performed phenomenological analysis in comparison with available experimental results for quantum dots made of CdSe and observe a very good agreement in this regard. Our experimentally consistent theoretical results also help in mapping the probability density of electron and hole inside spherical quantum dot. The consistency of our results with available experimental data signifies the capability as well as applicability of the ansatz for the effective confining potential to have reasonable information in the study of real nano-structured spherical systems.


\end{abstract}
\end{center}


\section{Introduction}
\label{intro}

The pursuit of physics in less than three dimensions is not only a mean to satisfy instinctive curiosity but also a principal tool for the advancement and progress of civilization. The last few decades have been witnessed the revolution in our civilization made by science and technology associated with nano-structured 2D (thin film), 1D (nano wire) and 0D (quantum dot)materials (see for example Ref. \cite{REV1,REV2,REV3,REV4,EMA2,QD1,QD2,QD3,QD4,QD5}).  Among the available low dimensional structures, Quantum Dots(QD), the most functional and reproducible nano-structures available to researchers have attracted much attention recently with their unique optical properties towards significant promises for a wide range of optoelectronic devices, including solar cells, photodetectors, and lasers.\cite{QD1,QD2,QD3,QD4,QD5,APP} The underlying phenomena behind the intriguing and valuable attributes of quantum dot is Quantum Confinement and hence investigation on the origin and effect of confinement in QD have become one of the most essential and exciting forefront fields in modern sciences.\cite{EMA2,QC1,QC2,QC3,QC4,QC5}

\ There are several theoretical as well as experimental efforts towards proper understanding of confinement mechanism in QDs. Theoretically, a wide verity of approaches have been employed to elucidate this phenomena. They include the tight-binding method,\cite{TBM1,TBM2,PEREZ,LIPLAN} the effective bond-orbital model,\cite{ebom} the effective mass approximation,\cite{EMA1,EMA2,EMA3} multi-band envelope-function models\cite{MBEFM} etc. However all of them have more or less limitations either in the accuracy of the predicted observable associated with confinement phenomena or in the computation cost. Although these methods are widely applied to investigate the structural, electronic and optical properties of nanostructures, still they are not quite straight forward; on the other hand the effective mass approximation (EMA) method has maintained a good track record over the last two decades in terms of accuracy, reliability, and efficiency. Within effective mass approximation the single particle (electron or hole) confinement is governed by the Schrodinger’s equation\cite{EMA2,EMA1,EMA3}

\begin{eqnarray}
H\Psi= E\Psi
\end{eqnarray}

\noindent and the hamiltonian of an electron or hole with effective mass $m^*$ trapped in a QD  is expressed by

\begin{eqnarray}
H=-\frac{\hbar^2}{2m^*}\nabla^2 + V(r),
\end{eqnarray}

\noindent which requires proper definition of potential function $V(r)$ including all the effective potentials associated with confining electrons or holes in QD. Remarkable theoretical efforts, in combination with many successful experimental programmes have been devoted to identify the realistic profile of the confining potentials. In literature there are essentially two main classes of approaches in order to obtain confining potentials: one uses first principles\cite{FP1,FP2,FP3,FP4} and others assume parameterize form, which allows us to model its shape. Taking into account the advantage of being greater flexibility, recent studies have been observed to pay significant deliberation towards the latter one with respect to the former and as a result considerable progress is observed in the investigation of quantum confinement by means of speculating different reliable theoretical models, which includes the rectangular potential well,\cite{RP1} harmonic oscillator potential,\cite{HOP1} Gaussian potential,\cite{GP1} Coulomb potential,\cite{CP1}  Rosen-Morese potential, \cite{RMP} Poschl-Teller potential,\cite{PTP1} Tietz potential,\cite{TP} Woods - Saxon potential,\cite{WSP1} Ellipsoidal potential, \cite{EP1}  Manning-Rosen (M-R) potential,\cite{MRP} Hulthen potential,\cite{HP1} etc.

\ In addition to these, a few investigative calculations have also been reported earlier based on the combined potential, $V (r) = ar^2 - \frac{b}{r}+ cr$  comprising of harmonic, linear and Coulomb terms as confinement potential.\cite{CPH1,CPH2,CPH3,CPH4,CPH5,CPH6} Usually this type of potential, known as Cornell plus Harmonic (CpH) confining potential is employed to study the bound states of the quarks in mesonic $(q\overline{q})$ system.\cite{meson} Here, the advantage of conjoined harmonic oscillator and coulomb potential is that it offers an impenetrable spherical cavity, which also includes the one gluon exchange Coulomb like potential between quarks within Quantum Chromodynamics (QCD) formalism and the contribution of linear term is due to the expectation from color confinement. In Ref. \cite{CPH1,CPH2,CPH3,CPH4,CPH5,CPH6}, the spherical quantum dot consisting an electron-hole pair was considered to be analogous to the mesonic system consisting of a $q - \overline{q}$ pair and hence their confinement mechanism is also similar. However, as far spherical quantum dot is concerned, where no such color interaction is present, the contribution due to linear term can be successfully excluded from the confining potential model. As a result we may expect $V (r) = ar^2 -\frac{b}{r}$ as the effective potential associated with confining electrons and holes within QD.

\ In this paper, considering confinement of an electron or hole under the conjoined harmonic oscillator and coulomb potential of the form $V (r) = ar^2 -\frac{b}{r}$, along with an  appropriate boundary condition we have solved the corresponding Schrodinger equation governed by EMA for ground state energy and wave function of the electron and hole and determined the shift in energy of conduction band minimum(CBM) and valence band maximum(VBM) due to quantum confinement. Using these quantum confinement induced shifts we have determined the size dependency of the band gap of semiconductor. In order to verify our theoretical predictions on confinement induced shift in Energy of CBM and VBM as well as the band gap in spherical quantum dot, we have performed phenomenological analysis in comparison with available experimental results for quantum dots made of CdSe. In addition, utilising these phenomenologically successful results we have mapped the probability density distribution inside CdSe spherical quantum dot. We have chosen CdSe here as the model systems only because of the existence of wealthy collection of experimental data\cite{QC2,data} for them, however we expect that our formalism can be extended to incorporate other semiconductor systems also.

\section{Theoretical framework}

Within Effective Mass Approximation (EMA), we define the hamiltonian for an electron (with effective mass $m_e^*$) confined in a spherical quantum dot as

\begin{eqnarray}
H=-\frac{\hbar^2}{2m_e^*}\nabla^2 + V(r), \hspace {20mm} V(r)=ar^2-\frac{b}{r},
\label{ham}
\end{eqnarray}

\noindent where $a$ and $b$ are the coupling parameters and assume that $a>0$ and the coulomb term is attractive, i.e., $b>0$. For this hamiltonian corresponding Schrodinger equation in three dimensions can be written as

\begin{eqnarray}
-\frac{\hbar^2}{2m_e^*}\bigg[\frac{\partial^2}{\partial r^2}+\frac{2}{r}\frac{\partial}{\partial r}-\frac{L^2(\theta,\phi)}{\hbar^2r^2}\bigg]\Psi(r,\theta,\phi)+V(r)\Psi(r,\theta,\phi)=E\Psi(r,\theta,\phi).\label{SE1}
 \end{eqnarray}

\noindent For spherically symmetric potential the eigen function $\Psi(r,\theta,\phi)=R_{nl}(r)Y_l^m(\theta,\phi)$  consists of the radial part $R_{nl}(r)$ and the spherical harmonics, $Y_l^m(\theta,\phi)$ representing the angular part which satisfies $ L^2(\theta,\phi)Y_l^m(\theta,\phi)=l(l+1)\hbar^2Y_l^m(\theta,\phi)$,
 with $l=0,1,2,3..... etc.$, the angular momentum quantum number. On the other hand, the radial function, to be well behaved must satisfy $R_{nl}(\infty)\rightarrow 0$, which in turn lead us to consider $ R_{nl}(r)=N_{nl} \frac{\psi_{nl}(r)}{r}$, including the normalization constant $N_{nl}$. Performing a little mathematical simplification the three dimensional Schrodinger equation for spherically symmetric potential $V(r)$ can be reduced to one dimensional equation  involving only the radial part and in terms of $\psi_{nl}(r)$ one can express it as

\begin{eqnarray}
\frac{d^2\psi_{nl}(r)}{d r^2}+\bigg[\Sigma_{nl}-Ar^2+\frac{B}{r}-\frac{l(l+1)}{r^2}\bigg]\psi_{nl}(r)=0,\label{rse}
\end{eqnarray}

\noindent with $\Sigma_{nl}=\frac{2m_e^*}{\hbar^2} E$, $A=\frac{2m_e^*}{\hbar^2} a$ and $B=\frac{2m_e^*}{\hbar^2}b$.

\ In order to solve this equation we consider the following ansatz as the trial wave function\cite{TWF,CPH2,CPH3}:

\begin{eqnarray}
\psi_{nl}(r)=f_n(r)g_l(r)= r^{l+1}Exp\bigg[-\alpha r^2-\beta r\bigg]f_n(r),\label{sol}
\end{eqnarray}

\noindent with

\begin{eqnarray}
f_n(r)=\left\{
           \begin{array}{ll}
             1, & \hbox{n=0;} \\
             \prod_{i=1}^{n}\bigg(r-\alpha_i^{(n)}\bigg), & \hbox{n=1,2,3....,}
           \end{array}
         \right.
\end{eqnarray}

\begin{eqnarray}
g_l(r)=r^{l+1}Exp\bigg[-\alpha r^2-\beta r\bigg].
\end{eqnarray}

\noindent On substitution of Eq.(\ref{sol}) in Eq. (\ref{rse}) we obtain

\begin{eqnarray}
rf_n''(r)+\bigg[-4\alpha r^2-2\beta r+2(l+1)\bigg]f_n'+\bigg[(4\alpha^2-A)r^3+4A\beta r^2+\bigg\{-2\alpha(l+3)+\beta^2+\Sigma\bigg\}r\nonumber\\+\bigg\{B-2\beta(l+1)\bigg\}\bigg]f_n(r)=0.\label{gense}
\end{eqnarray}

\noindent Comparing above equation with

\begin{eqnarray}
\Bigg(\sum_{i=0}^{k+2}\mu_{k+2,i} x^{k+2-i}\Bigg)y''+\Bigg(\sum_{i=0}^{k+1}\nu_{k+1,i} x^{k+1-i}\Bigg)y'-\Bigg(\sum_{i=0}^{k}\tau_{k,i} x^{k-i}\Bigg)y=0
\end{eqnarray}

\noindent and making use of the Theorem 6 of Ref. \cite{THEOREM}, which suggests the following necessary condition to have polynomial solution of degree $n$:

\begin{eqnarray}
\tau_{k,0}=n(n-1)\mu_{k+2,0}+n\nu_{k+1,0}, \hspace{15mm} k=0,1,2,3,.....\label{theorem},
\end{eqnarray}

\noindent we obtain

\begin{eqnarray}
-2\alpha(l+3)+\beta^2+\Sigma_{nl}=4n\alpha,
\end{eqnarray}

\noindent which in turn lead to have

\begin{eqnarray}
\Sigma_{nl} = 2\alpha(2n+l+3)-\beta^2
\label{sigma}
\end{eqnarray}

\noindent Moreover, substituting (\ref{sigma}) in (\ref{gense}) and then for the case $n=0$, we obtain following relations by comparing coefficients of like powers of $r$:

\begin{eqnarray}
\alpha=\frac{\sqrt{A}}{2}=\sqrt{\frac{2m_e^*}{\hbar^2}}\frac{\sqrt{a}}{2},
\end{eqnarray}

\begin{eqnarray}
 \beta = 0,
\end{eqnarray}

\noindent and

\begin{eqnarray}
 B = 0.
\end{eqnarray}

\noindent These relations, along with (\ref{sigma}) and (\ref{sol}), lead us to have the ground state energy eigen values and radial eigen function for an electron with effective mass $m_e^*$  as

\begin{eqnarray}
E^e_{0l}=\sqrt{\frac{\hbar^2}{2m_e^*}}\sqrt{a}(l+3)
\label{eenergy}
\end{eqnarray}

\noindent and

\begin{eqnarray}
 R^e_{0l}(r)=N_{0l}\frac{\psi_{0l}(r)}{r}=\frac{1}{r}N_{0l}r^{l+1}Exp\bigg[-\sqrt{\frac{2m_e^*}{\hbar^2}}\frac{\sqrt{a}}{2} r^2\bigg]
\label{ewf}
\end{eqnarray}

\noindent respectively. Here $N_{0l}$ is the normalisation constant for $n=0$.
\ In a similar way, the ground state energy eigen value and eigen function for a hole with effective mass ``$m_h^*$" can be expressed as

\begin{eqnarray}
E^h_{0l}=\sqrt{\frac{\hbar^2}{2m_h^*}}\sqrt{a}(l+3)
\label{henergy}
\end{eqnarray}
\noindent and
\begin{eqnarray}
 R^h_{0l}(r)=\frac{\psi_{0l}(r)}{r}=\frac{1}{r}N_{0l}r^{l+1}Exp\bigg[-\sqrt{\frac{2m_h^*}{\hbar^2}}\frac{\sqrt{a}}{2} r^2\bigg]
\label{hwf}
\end{eqnarray}
\noindent respectively. Further the same formalism can be extended in order to account for the energy states corresponding $n=1,2,3,...etc.$, however here we are interested on investigating ground state properties of quantum dot and thus we restrict ourself with $n=0$  and do not concern about the higher energy states.

\section{Determination of the parameter space}

\ It is observed that the expressions for  the ground state energy and wavefunction of electron involve the parameters $N_{0l}$ and $a$ to be determined. In general the theoretical works\cite{CPH2,CPH4,CPH5,CPH6} that deal with solution of Schrodinger equation with polynomial confining potentials for spherical quantum dot determine the parameter space either numerically using suitable constraints or use the values for the parameters obtained by Chaudhuri and Mondal\cite{CPH1} with supersymmetric quantum mechanics. However as discussed in Ref. \cite{CPH3}, most of them did not do any attempt to verify the validity of the model/parameters by contrasting it with actual experimental data. In this paper, in order to determine the parameter space, we utilize the relations for normalization condition and the behaviour of wave function and probability density at $r=R$ and the acceptable range of the parameters are identified using available experimental information.  In this regard, firstly we consider the boundary condition that the conjoined potential adheres an electron inside an impenetrable spherical QD of radius $R$, i.e, probability of finding the particle beyond $R$ is zero or $\int_0^R |R^e_{0l}(r)|^2dr=1$ and then we obtain the normalization constant in terms of $a$ as

\begin{eqnarray}
|N_{0l}|^2=\frac{1}{\int_0^R \bigg|r^{l}Exp\bigg[-\sqrt{\frac{2m_{e}^*}{\hbar^2}}\frac{\sqrt{a}}{2} r^2\bigg]\bigg|^2dr},
\label{norm}
\end{eqnarray}

\noindent which in turn lead us to have the wave functions as well as the probability densities as

 \begin{eqnarray}
 R^{e}_{0l}(r)=\sqrt{\frac{1}{\int_0^R \bigg|r^{l}Exp\bigg[-\sqrt{\frac{2m_{e}^*}{\hbar^2}}\frac{\sqrt{a}}{2} r^2\bigg]\bigg|^2dr}}r^{l}Exp\bigg[-\sqrt{\frac{2m_{e}^*}{\hbar^2}}\frac{\sqrt{a}}{2} r^2\bigg]
\end{eqnarray}

\noindent and

 \begin{eqnarray}
 \bigg|R^{e}_{0l}(r)\bigg|^2=\frac{\bigg|r^{l}Exp\bigg[-\sqrt{\frac{2m_{e}^*}{\hbar^2}}\frac{\sqrt{a}}{2} r^2\bigg]\bigg|^2}{\int_0^R \bigg|r^{l}Exp\bigg[-\sqrt{\frac{2m_{e}^*}{\hbar^2}}\frac{\sqrt{a}}{2} r^2\bigg]\bigg|^2dr}
\end{eqnarray}

\noindent respectively. The nature of the wave function and incorporated boundary condition require that both the wave function and probability density must vanish at $r=R$ and beyond for a spherical quantum dot of radius $R$, which in turn requires $a=\infty$ and lead to the physically nonsense result, $E^e_{0l}=\infty$. Thus under this consideration, determination of physically acceptable value for $a$ by means of solving  $R^e_{0l}(r)\bigg|_{r=R}=0$ and $\bigg|R^{e,h}_{0l}(r)\bigg|_{r=R}^2=0$ is not possible. However leaving a little scope for penetration, instead of being considering probability density to be exactly zero at $r=R$, we may expect suitable information about $a$. With this expectation, we now consider the total probability, $\int_0^R |R_{0l}(r)|^2dr =x=0.999$ and hence the probability density becomes
 \begin{eqnarray}
 \bigg|R^{e}_{0l}(r)\bigg|^2=\frac{x\bigg|r^{l}Exp\bigg[-\sqrt{\frac{2m_{e}^*}{\hbar^2}}\frac{\sqrt{a}}{2} r^2\bigg]\bigg|^2}{\int_0^R \bigg|r^{l}Exp\bigg[-\sqrt{\frac{2m_{e}^*}{\hbar^2}}\frac{\sqrt{a}}{2} r^2\bigg]\bigg|^2dr}
\label{pd}
\end{eqnarray}

\begin{figure*}
\centering
\includegraphics[scale=0.8]{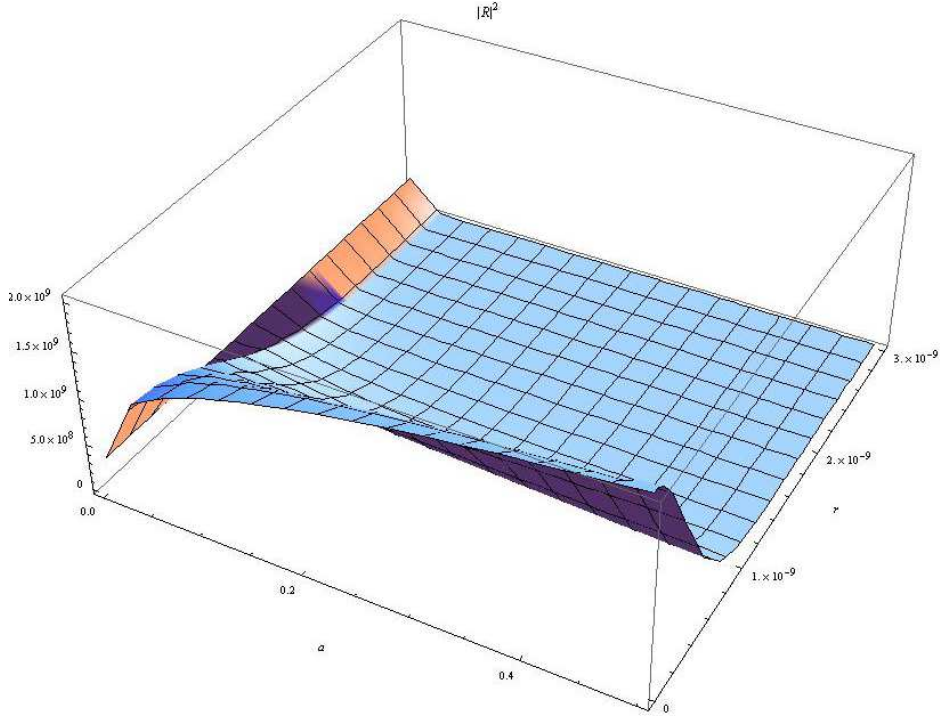}
\caption{ Radial probability density of an electron within a spherical quantum dot of radius $3 nm$ as a function of the parameter $a$ and $r$.}
\label{density}
\end{figure*}

\begin{figure*}
\centering
\includegraphics[scale=0.5]{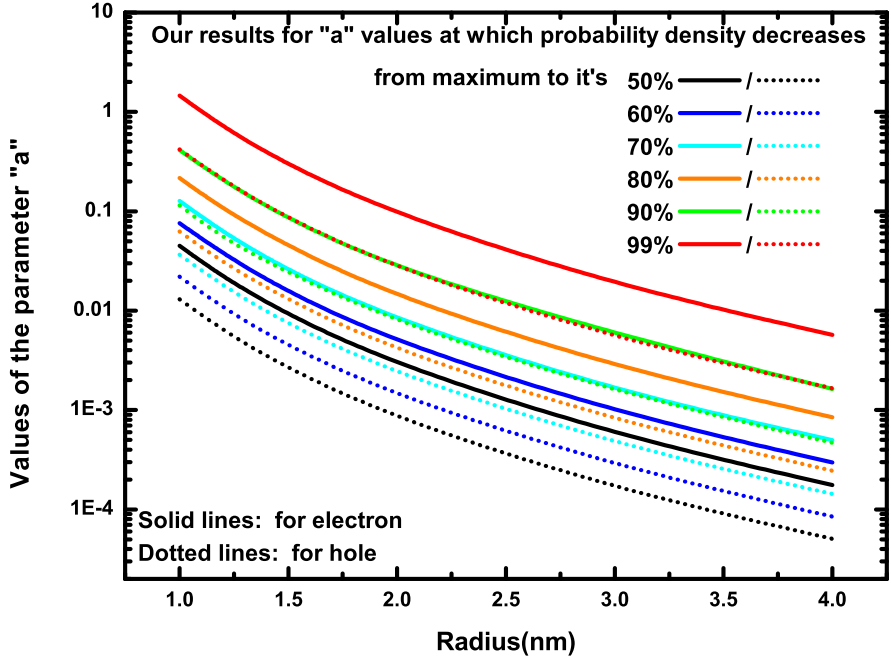}
\caption{Dependence of the parameter $a$ for quantum dots of different radius.}
\label{aReh}
\end{figure*}

\noindent In Fig. \ref{density}, the probability density for an electron with effective mass $m_e^*=0.13 m_e$ within a CdSe QD of radius $R=3 nm$ is shown as a function of the parameter $a$ and position $r$. As far variation of probability density with respect to $r$ for fixed $a$ is concerned, we observe a constant value $\frac{x}{R}$  at $a=0$ for $\bigg|R^{e}_{0l}(r)\bigg|_{r=R}^2$ throughout the region within QD and for other non-zero $a$ values it takes a maximum value $N_{0l}$ at $r=0$ and decreases towards the surface of the QD.    Now, if we concentrate on the variation of probability density with respect to $a$, an increase in $\bigg|R(a,r)\bigg|^2$  with $a$ is observed in the central region of the QD and as we move radially outward, $\bigg|R(a,r)\bigg|^2$ decreases as $a$ increases. Further at $r=R$, we observe that $\bigg|R^{e}_{0l}(r)\bigg|_{r=R}^2$ has the maximum possible value $\frac{x}{R}$ for $a=0$ and decreases with increase in $a$. We also estimate that probability density at $r=R=3 nm$ becomes $50\%$, $60\%$, $70\%$, $80\%$, $90\%$ and $99\%$ less of its initial value, i.e., becomes $\frac{0.5x}{R}$, $\frac{0.4x}{R}$, $\frac{0.3x}{R}$, $\frac{0.2x}{R}$, $\frac{0.1x}{R}$ and $\frac{0.01x}{R}$ respectively for $a=0.0005619$, $a=0.0009465$, $a=0.00155$, $a=0.0026831$, $a=0.00581$ and $a=0.0180731$. Further, it is well known that the total probability within the QD doesn't depend on $a$, instead it controls the probability of finding the electron within it's different regions. Larger the value of $a$ signifies greater possibility of finding the electron within central region and accordingly to have the electron beyond central region $a$ needs to decrease. On the other hand, in accord with our presumption $a>0$( see Eq. (\ref{ham})), it is confirm that the probability of finding electron within QD cannot be uniform throughout the region and hence $\bigg|R^{e}_{0l}(r)\bigg|_{r=R}^2 < \frac{x}{R}$. A data driven formalism may lead to confirmative picture of probability distribution of the electron inside QD, however in this paper instead of looking for accurate probability density of the electron leading to definite $a$ value, firstly we neglect the possibility $\bigg|R^{e}_{0l}(r)\bigg|_{r=R}^2 \ll\frac{x}{R}$ and proceed our study considering a series of distribution of electron such that the probability density takes the values $\frac{0.5x}{R}$, $\frac{0.4x}{R}$, $\frac{0.3x}{R}$, $\frac{0.2x}{R}$, $\frac{0.1x}{R}$ and $\frac{0.01x}{R}$ at $r=R$, i.e., it decreases by $50\%$, $60\%$, $70\%$, $80\%$, $90\%$ and $99\%$ respectively on the surface from its maximum possible value $\frac{x}{R}$(which depends on $R$ and occurs at $a=0$). We have estimated the corresponding $a$ values for which these series of distributions occur for quantum dots of different radius and plotted them in Fig. \ref{aReh} against radius $R$ of CdSe quantum dots. In a similar way, we have also determined the corresponding values for the parameter $a$, responsible for these series of densities of hole in CdSe QDs and shown in Fig. \ref{aReh} by dotted lines. In this case and in calculations throughout the paper, we have used the effective mass of electron and hole are considered to be $m_e^*=0.13 m_e$ and $m_h^*=0.45 m_e$ respectively. In addition, we would also like to mention that all the computations, in this paper are performed by using MATHEMATICA 9.\cite{MATHEMATICA}

\section{Results and discussion}

\begin{figure*}
\centering
\includegraphics[scale=0.5]{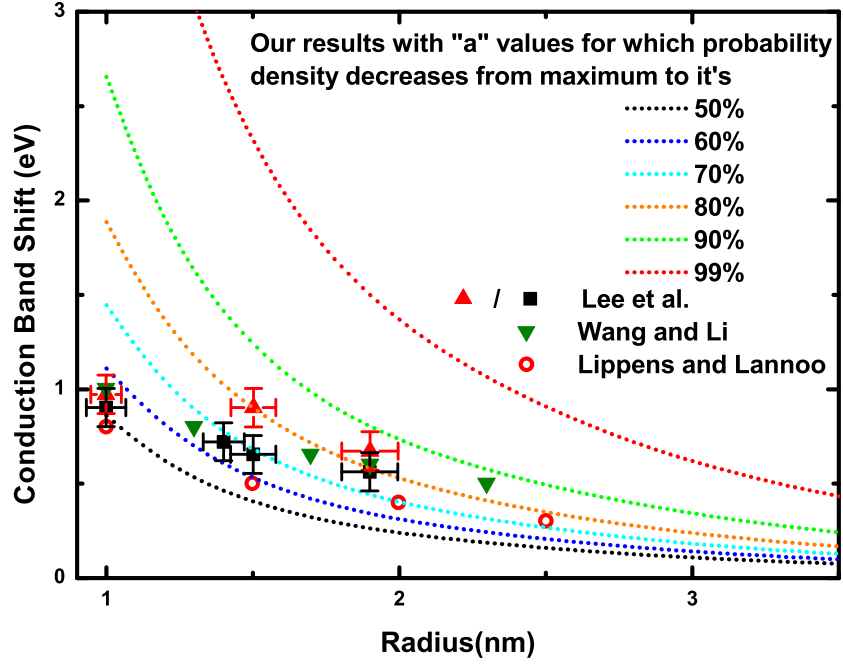}
\caption{Quantum confinement induced shift in energy of minimum of conduction band (CBM) of spherical quantum dots.}
\label{ECDSER}
\end{figure*}

\begin{figure*}
\centering
\includegraphics[scale=0.5]{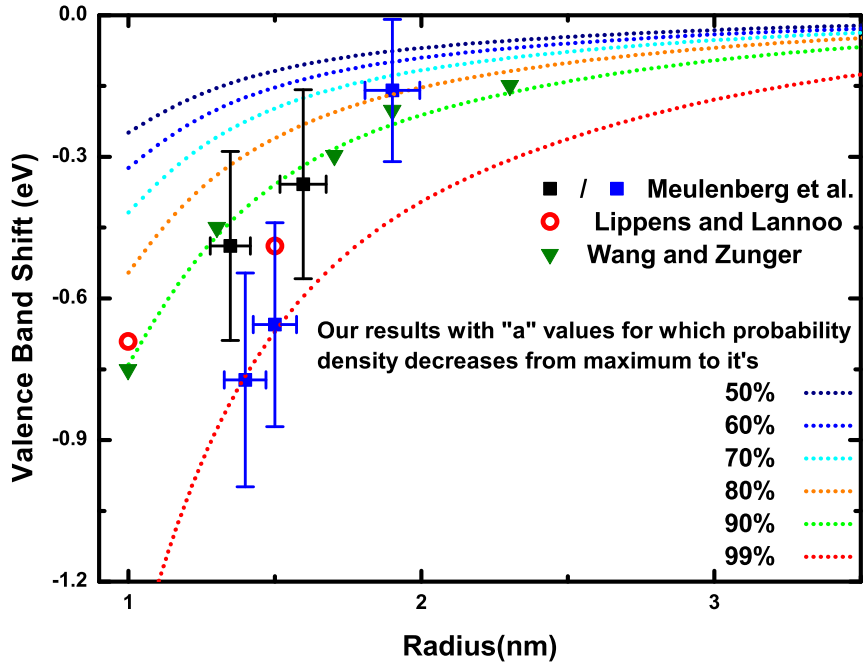}
\caption{Quantum confinement induced shift in energy of maximum of valence band (VBM) of spherical quantum dots.}
\label{HCDSER}
\end{figure*}

The equations (\ref{eenergy}) and (\ref{henergy}) provide the shift in ground state energy of electron and hole or  the energies of CBM and VBM due to quantum confinement relative to the corresponding bulk energy values. We have computed the ground state energy shifts in accord with Eqs.(\ref{eenergy}) and (\ref{henergy}) using above mentioned $a$ values for which probability density decreases from maximum to it's $50\%$, $60\%$, $70\%$, $80\%$, $90\%$ and $99\%$ on the surface of CdSe quantum dot, and corresponding shifts are shown in Fig. \ref{ECDSER} and Fig. \ref{HCDSER} as a function of their radius($R$). Our theoretical results for CBM and VBM for the six possible probability densities of electron and hole are represented by colored dotted lines and accordingly indicated in the respective figure. In Fig. \ref{ECDSER}, we have included the experimental data points obtained in Ref. \cite{PRLLEE} for CBM using x-ray absorption spectroscopy (XAS) technique, along with the theoretical results obtained by Wang and Li\cite{WANGLI} and Lippens and Lannoo\cite{LIPLAN}. The studies on band edge shift in Ref.\cite{WANGLI} is based on based the charge patching (CP) theory, which uses charge density instead of local potential to have electronic structure in quantum dot, while  Lippens and Lannoo\cite{LIPLAN} used tight binding (TB) approach to study effect of confinement on band edge energy. Similarly the experimental data points of Ref. \cite{ACSMEU} using x-ray absorption and photoemission spectroscopy and the theoretical predictions by Wang and Zunger\cite{WANGZUNG} as well as Lippens and Lannoo\cite{LIPLAN} have also been plotted along with our theoretical results for VBM shift in Fig. \ref{HCDSER}.

\begin{figure*}
\centering
\includegraphics[scale=0.5]{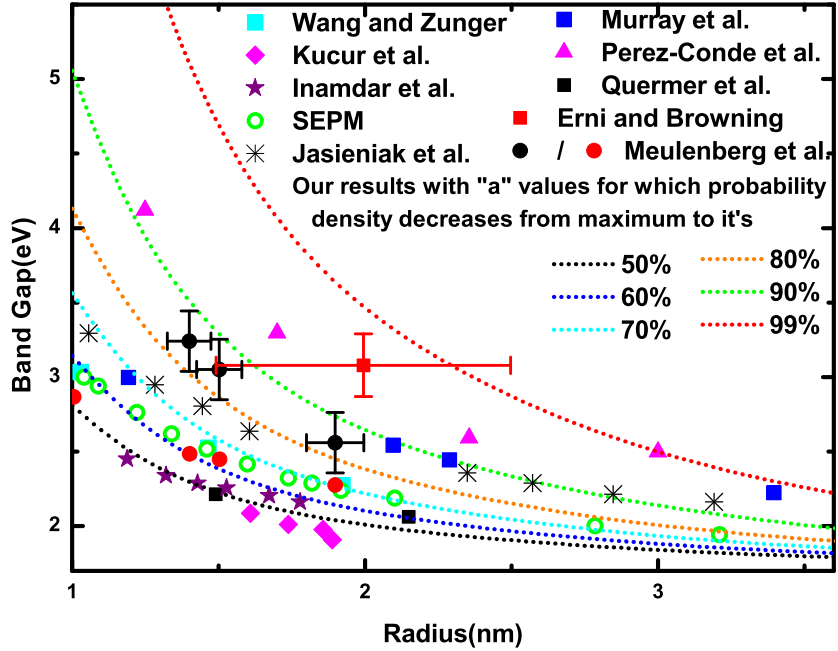}
\caption{Size dependency of band gap energy of spherical quantum dots.}
\label{EGCDSER}
\end{figure*}

\ The band gap for CdSe QD is calculated using $E_g^{QD}=E_g^{Bulk}+E_{00}^e+E_{00}^h$ and the corresponding results for size dependency of band gap are displayed in Fig. \ref{EGCDSER}. Here our results for the series of probability densities mentioned above are represented by the colored dotted lines as indicated in the figure. Alongside our results, we have plotted a large number of other available experimental and theoretical results obtained by Wang and Zunger,\cite{WANGZUNG} Kucur et al.,\cite{KUCUR} Inamdar et al.,\cite{INAMDAR} Querner et al.,\cite{QUERNER} Erni and Browning,\cite{ULTRAROLF} Murray et al.,\cite{MURRAY} Meulenberg et al.,\cite{ACSMEU} Jasieniak et al.,\cite{JASIENIAK} Perez-Conde and Bhattacharjee\cite{PEREZ} and SEPM \cite{JASIENIAK}. The theoretical results reported by Jasieniak et al., \cite{JASIENIAK} are based on the semi-empirical pseudo-potential method(SEPM), while the experimental measurements by Erni and Browning\cite{ULTRAROLF} utilised Valence electron energy-loss spectroscopy (VEELS) and Murray et al.\cite{MURRAY} used optical absorption spectroscopy. On the other hand, the experimental works on quantum confinement by Kucur et al.,\cite{KUCUR} Inamdar et al.,\cite{INAMDAR} and Querner et al.,\cite{QUERNER} were performed through cyclic voltammetry method and Jasieniak et al.\cite{JASIENIAK} investigated the size dependency of valence and conduction band-edge energies using photoelectron spectroscopy in air (PESA).

\begin{figure*}
\centering
\includegraphics[scale=0.8]{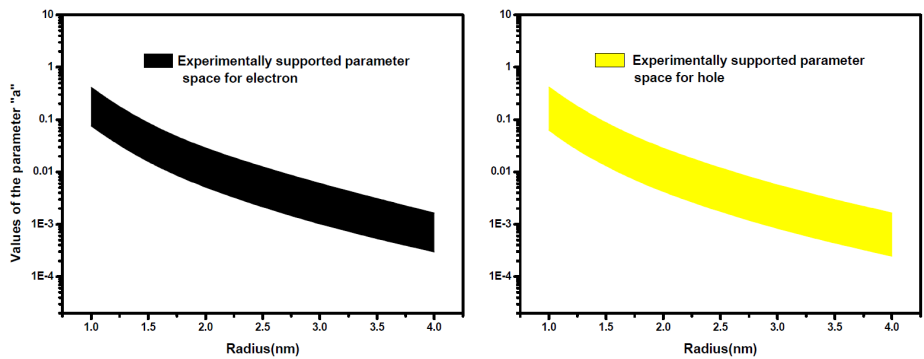}
\caption{Experimentally supported parameter spaces for distribution of electron and hole against radius of spherical quantum dot.}
\label{PSPACE}
\end{figure*}

\begin{figure*}
\centering
\includegraphics[scale=0.5]{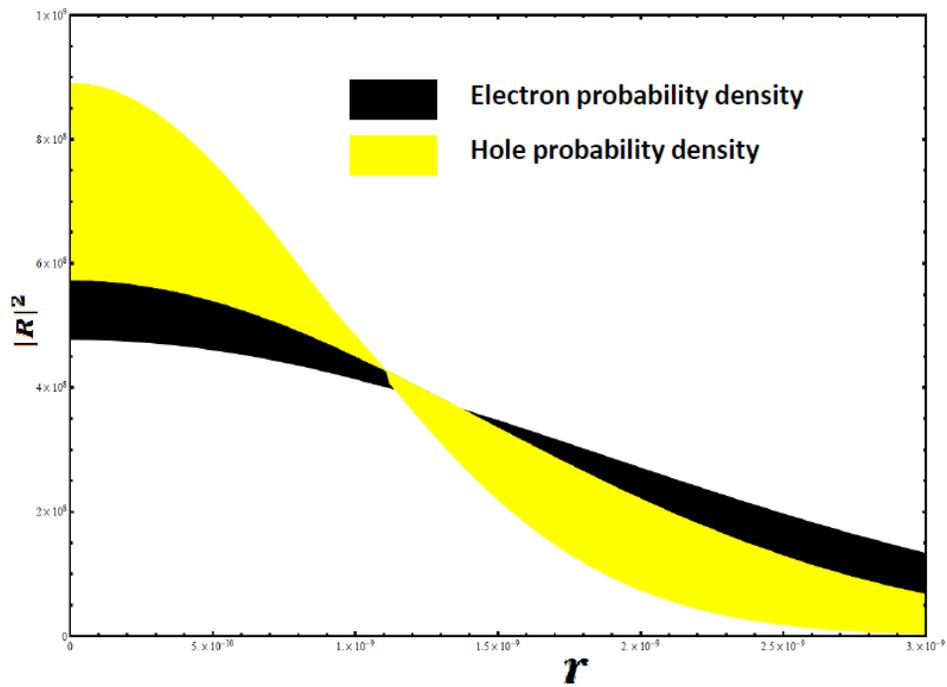}
\caption{Experimentally supported radial probability density of an electron and a hole within a spherical quantum dot of radius $3 nm$.}
\label{PROBDEN}
\end{figure*}

\ Electrons and holes in bulk semiconductor observe allowed and forbidden energy bands. However in low dimensions, as a result of quantum confinement they observe discrete energy states instead of bands, the energy values of which varies with size of the QD.  Individual shift in CBM and VBM are depicted in Fig. \ref{ECDSER} and  \ref{HCDSER} with respect to corresponding values of bulk CdSe. Quantum confinement induced shift in band gap is also shown in Fig. \ref{EGCDSER} as a function of size of the QDs. In addition to available theoretical results, for the purpose of direct comparison, several experimental data have been plotted alongside our results. From the respective figures, we observe that both the conduction band and valence band edges of the QD are shifted to higher or lower energies relative to their bulk counterpart. From Fig. \ref{ECDSER}, it is observed that most of the experimental as well as theoretical results for conduction band energy shift lie in between our results obtained with the $a$ values for which probability density of the electron decreases by $60\%$ and $90\%$ from it's maximum on the surface. On the other hand, experimental as well as theoretical results for valence band shift depicted in Fig. \ref{HCDSER} lie within the two curves obtained with $a$ values for which probability density of the hole becomes $80\%$ and $99\%$ less on the surface of the quantum dot. It is also observed that Wang and Zunger\cite{WANGZUNG} results for VBM are well explicable with our results obtained with $a$ values for which probability density of the hole becomes $80\%$ of its maximum on the surface. Further, Fig. \ref{EGCDSER} reflects that most of the experimental and theoretical data reside in between our results obtained by considering $a$ values for which the probability density of electron and hole becomes $90\%$ and $50\%$ less from the maximum possible value $\frac{x}{R}$ on the surface.

\ The experimental data points, measured on the basis of several methods, are observed to scatter within a wide range of radius $R$. Therefore from the comparative analysis with such data points it is very difficult for us to identify a definite and well consistent track for the quantum confinement induced deviation of CBM, VBM and band gap energy from the corresponding bulk values with respect to size of the spherical quantum dot and hence we can neither estimate $a$ values accurately nor identify the most possible probability density distribution of electron and hole within spherical quantum dot. Instead we are capable of predicting a range of parameter space $a$ for which overall consistency of our theoretical results with other available experimental data may be achieved and in addition, the corresponding acceptable region of probability density distribution of electron and hole can be determined. Based on the analysis of our results depicted in Fig.\ref{ECDSER} and Fig. \ref{HCDSER} in comparison with several experimental data we identify that acceptable range of $a$ values lie within the two curves for which the probability density decreases upto $60\%$ and $90\%$ for electron and $80\%$ and $99\%$ for hole on the surface of the quantum dot from the maximum possible value and it is shown in Fig.\ref{PSPACE}. In Fig. \ref{PROBDEN}, we present the the probability density distribution of electron and hole within a $3 nm$ CdSe quantum dot which provides well description of available experimental data through the Schrodinger equation under effective mass approximation method and with $V(r)=ar^2-\frac{b}{r}$ as the effective confining potential. Further, to the best of our knowledge, since there is no conclusive experimental information about the probability density distribution of electron and hole within quantum dot, comparative analysis in this regard is not possible and expect that future experimental work will provide reliable information in this regard and help us to verify the parameter space and lead to have more accurate and precise prediction on confinement induced shift in energy of CBM, VBM and band gap of spherical quantum dot.

\ Although the chosen potential model under effective mass approximation along with above mentioned range of values for the parameter $a$, provide considerable consistency with experimental data  but better realization of consistency is within $3 nm - 1nm$, i.e., at larger particle size, and observe that our results diverge from experimental observations for smaller ($\leq 1 nm$) quantum dots. Available experimental data initially experience quantum confinement and hence induced shift increase with decrease in size, however towards $1 nm$ the confinement induction reduce and pinning of the conduction band edge occur with an average shift of $\sim 1 eV$ from the bulk values. This contradiction between theoretical expectation and experimental observation, in accord with discussion in Ref. \cite{PRLLEE} is due to the fact that there are energy states in the upper conduction band of QDs that are inherently unaffected by quantum confinement, unlike the states at the bottom of the CB which are strongly affected. When size decreases it is expected that the low energy states under confinement effect shift to such a degree that the bottom of the conduction band is now dominated by the higher energy states without experiencing quantum confinement effect. On the other hand, Puzder et al.\cite{PUZDER} suggested that the pinning of conduction band edge may be due to surface states, which are different from bulk states. At this end, we expect to extend our formalism including such effects to incorporate properties of smaller QD in a future communication.

\section{Conclusion}

Identification of effective interaction potential functions for the real systems is of fundamental importance in different branches of modern science because it provides the most efficient way to summarize what we know about a complex system. In the field of studying quantum dots also we observe existence of a wide variety of potential functions in Ref. \cite{RP1,HOP1,GP1,CP1,RMP,PTP1,TP,WSP1,EP1,MRP,HP1,CPH1,CPH2,CPH4,CPH5,CPH6} and some of them achieved significant phenomenological success. However most of them neither performed phenomenological analysis of their results nor  paid  attention to validate their models by contrasting it with actual experimental data. In this paper we have proposed an ansatz of the form $V(r)=ar^2-\frac{b}{r}$, combining harmonic and coulomb interaction as the effective potential associated with confining electron/hole within spherical quantum dot and  using this ansatz along with suitable boundary conditions we have solved the corresponding Schrodinger equation governed by EMA for ground state energy and wave function of the electron and hole and determined the shift in energy of CBM, VBM and band gap of spherical quantum dot due to quantum confinement. We have also performed a phenomenological analysis of our theoretical prediction with several experimental results for CdSe QD systems. From the phenomenological analysis we have observed that the ansatz for effective potential $V(r)=ar^2-\frac{b}{r}$, with a specific range of $a$ values, within effective mass approximation provides a very good description of the energy shift of CBM and VBM due to confinement, which are consistent with other results taken from Ref.\cite{PRLLEE,WANGLI,LIPLAN,ACSMEU,WANGZUNG}. We have also achieved considerable consistency as far our results for size dependency of band gap energy for spherical QD are concerned in comparison with the experimental data taken from Ref. \cite{PEREZ,ULTRAROLF,MURRAY,KUCUR,INAMDAR,QUERNER,JASIENIAK} and other theoretically reported values in Ref.\cite{WANGZUNG,PEREZ,JASIENIAK}. The consistency achieved in this regard within the region of parameter space signifies the validity of our ansatz as the effective confining potential and the experimentally supported parameter space also helps us to predict the distribution of probability density of electron or hole within spherical quantum dots.

\ Our concluding impression based on all these observations is that the ansatz $V(r)=ar^2-\frac{b}{r}$ with conjoined harmonic oscillator and coulomb interaction as effective potential for confining electron or hole is capable of predicting successfully the quantum confinement induced shift in energy states as well as band gap of spherical quantum dot and confirm its candidature as a viable model to have reasonable information in the study of real nano-structured spherical systems.

\newpage

\end{document}